\documentclass[12pt]{iopart}
\usepackage{iopams}  

\usepackage{graphicx}
\usepackage{dcolumn}
\usepackage{bm}

\newcommand{\cm}{cm$^{-1}$} \newcommand{\A}{\AA$^{-1}$}

\begin{document}

\title[Macroscopic proton states] {A neutron diffraction study of macroscopically entangled proton states in the high temperature phase of the KHCO$_3$ crystal at 340 K}

\author{Fran\c{c}ois Fillaux}
\footnote[3]{To whom correspondence should be addressed:\\ fillaux@glvt-cnrs.fr\\ Tel: +33149781283\\ Fax: +33149781218 \\ (fillaux@glvt-cnrs.fr; http://www.ladir.cnrs.fr/pagefillaux\_eng.htm)}
\address{LADIR-CNRS, UMR 7075,
Universit\'{e} P. et M. Curie, 2 rue Henry Dunant, 94320 Thiais, France}
\author{Alain Cousson}
\address{Laboratoire L\'{e}on Brillouin (CEA-CNRS), C.E. Saclay, 91191
Gif-sur-Yvette, cedex, France}
\author{Matthias J Gutmann}
\address{ISIS Facility, Rutherford Appleton Laboratory, Chilton, Didcot,
OX11 0QX, UK}
\date{\today}

\begin{abstract}
We utilize single-crystal neutron diffraction to study the $C2/m$ structure of potassium hydrogen carbonate (KHCO$_3$) and macroscopic quantum entanglement above the phase transition at $T_c = 318$ K. Whereas split atom sites could be due to disorder, the diffraction pattern at 340 K evidences macroscopic proton states identical to those  previously observed below $T_c$ by F. Fillaux et al., (2006 \textit{J. Phys.: Condens. Matter} \textbf{18} 3229). We propose a theoretical framework for decoherence-free proton states and the calculated differential cross-section accords with observations. The structural transition occurs from one ordered $P2_1/a$ structure ($T < T_c$) to another ordered $C2/m$ structure. There is no breakdown of the quantum regime. It is suggested that the crystal is a macroscopic quantum object which can be represented by a state vector. Raman spectroscopy and quasi-elastic neutron scattering suggest that the $|C2/m\rangle$ state vector is a superposition of the state vectors for two $P2_1/a$-like structures symmetric with respect to $(a,c)$ planes. 
\end{abstract}

\pacs{ 61.12.-9, 03.65.Ud, 63.10.+a, 82.30.Hk.\\ Keywords: Quantum entanglement, Neutron diffraction, Hydrogen bonding, Proton tunneling, phase transition}

\maketitle

\section{Introduction}

In principle, a crystal with perfect translational invariance is a macroscopic quantum systems with discrete phonon states at any temperature below melting or decomposition. However, for crystals containing O--H$\cdots$O hydrogen bonds, there is a widely-spread dichotomy of interpretation for the so-called ``proton disorder'', usually observed as a dynamical equilibrium between two sites, say $\mathrm{O1-H}\cdots \mathrm{O2}$ and $\mathrm{O1}\cdots \mathrm{H-O2}$. For such systems, semiclassical approaches are commonly used to rationalize proton correlation times measured with solid-state NMR and quasi-elastic neutron scattering (QENS) \cite{ST}. Semiclassical protons are dimensionless particles, with definite positions and momentums, moving in a double-well coupled to an incoherent thermal bath. Dynamics are rationalized with uncorrelated jumps over the barrier and ``incoherent tunneling'' through the barrier. In fact, these models describe a liquid-like surroundings and strong interaction with the thermal bath is supposed to lead to fast decoherence. By contrast, diffraction and vibrational spectra provide unquestionable evidences that the translational invariance and the quantum nature of lattice dynamics are not destroyed by proton transfer. 

The crystal of potassium hydrogen carbonate, KHCO$_3$, is at variance with this dichotomy of interpretation \cite{IF,FCKeen,FCG2,Fil7}. This crystal is composed of centrosymmetric hydrogen bonded dimers (HCO$_3^-)_2$, separated by potassium ions. Below $\approx 150$ K, protons occupy sites corresponding to a unique configuration for dimers, say $L$. Between 150 and 300 K, protons are progressively transferred to the other centrosymmetric configuration $R$, up to a population ratio $R:L \approx 18:82$. In contrast to semiclassical models, it has been observed that the sublattice of protons is a macroscopically entangled quantum object over the whole temperature range. Quantum correlations outside any description offered by classical physics are evidenced with neutrons because the intensity diffracted by the sublattice is proportional to the total nuclear cross-section $\sigma_H \approx 82.02$ bn, instead of $\sigma_{Hc} \approx 1.76$ bn for uncorrelated protons. Enhanced diffraction observed from cryogenic to room temperatures is consistent with a superposition of decoherence-free macroscopic states at thermal equilibrium and the correlation time measured with QENS \cite{EGS} can be rationalized with quantum interferences \cite{Fil7}. There are therefore consistent evidences that protons in the crystal field are not semiclassical particles possessing properties on their own right \cite{Leggett}. 

In the present paper we examine whether the macroscopic entanglement is destroyed by the reversible structural phase transition at $T_c = 318$ K \cite{H}. The population ratio $R:L$ increases discontinuously to $0.5:0.5$ and then remains a constant above $T_c$. This transition, supposed to be of the order-disorder type \cite{KY,TTY,MYSIKF,KNHKKY1,KNHKKY2,KSNZKY,YK,LJ,LJ2}, could be of great significance to experimental studies of the quantum-to-classical interface in the crystalline state. However, structural and dynamical models are still in debate and require further studies. We have therefore conducted extensive single-crystal neutron diffraction measurements in order to (i) determine the best structural model; (ii) observe whether enhanced features due to quantum correlations survive or not above $T_c$. The experiments presented below show that the structural transition does not destroy the macroscopic quantum behaviour of protons. It transpires that the transition occurs from one ordered structure to another ordered structure through a coherent quantum process preserving the translational invariance of the crystal. 

The organization of this paper is as follows. In Sec. \ref{sec:2}, we determine the structure at 340 K, for the sake of comparison with previous models. In Sec. \ref{sec:3}, we examine the total differential cross-section measured over a large $\mathbf{Q}$-range, at the same temperature, which evidence the enhanced features of quantum correlations. The theoretical framework is presented in Sec. \ref{sec:4} and we calculate the differential cross-section of correlated protons in Sec. \ref{sec:5} for comparison with observations. In Sec. \ref{sec:6}, macroscopic states are opposed to statistical disorder and we utilize Raman spectroscopy to probe proton dynamics. We suggest that the crystal is in a pure state that can be represented by state vectors. 

\section{\label{sec:2}Crystal structure}

The reversible structural transition has a very small enthalpy and the crystal retains its optical transparency \cite{H}. Below $T_c$, the crystal symmetry is $P2_1/a$ with four indistinguishable KHCO$_3$ entities in the unit cell \cite{FCKeen,FCG2,TTO1,TTO2}. Planar centrosymmetric dimers (HCO$_3^-$)$_2$, with O$\cdots$O distances of $\approx 2.58$ \AA, are practically parallel to $(103)$ planes. They are alternately slightly tilted with respect to $(a,c)$ glide planes.  

Kashida and Yamamoto (KY) \cite{KY} have measured X-ray diffraction across the transition. The unit cell parameters are very little affected, the symmetry above $T_c$ is $C2/m$ and $(a,c)$ glide planes become mirror planes. These authors envisaged two structural models: (i) the ``displacive'' model based on reorientation of dimers whose symmetry changes from $C_i$ ($T \le T_c$) to $C_{2h}$ above $T_c$; (ii) the ``disorder'' model composed of split $C_i$ dimers, symmetric with respect to the mirror plane. This latter model was preferred by KY, but Machida et al. \cite{MYSIKF} preferred $C_{2h}$ dimers to fit their neutron diffraction data. 

We have collected single-crystal neutron diffraction data with the four-circle diffractometer 5C2 at the Orph\'{e}e reactor (Laboratoire L\'{e}on-Brillouin) \cite{LLB}. The twining-free crystal ($\approx 3\times 3\times 3$ mm$^3$), previously studied in Ref. \cite{FCG2} was loaded into an aluminum container heated at $(340 \pm 1)$ K. This temperature is sufficiently below decomposition ($\approx 400$ K) to avoid damage, and it is sufficiently high to marginalize critical phenomena \cite{EGS,H,KNHKKY1,KNHKKY2,KSNZKY}. The data reduction was carried out with CRYSTALS \cite{WPCBC,CRYSTALS}.

We confirm the monoclinic $C2/m$ ($C_{2h}^3$) symmetry and previous unit cell parameters (Table \ref{tab:1}) \cite{KY,MYSIKF}. Both structural models, referred to as ``$C_{2h}$'' and ``$C_i$'' in Table \ref{tab:1}, were refined. $C_{2h}$ dimers are composed of four O$_{1/2}$C$_{1/2}$OH$_{1/2}$ entities per unit cell, while $C_i$ dimers are obtained with four (HCO$_3$)$_{1/2}$ units. This latter model was refined in two steps. First, the split geometry was constrained to fit the mirror-plane symmetry. Then, constraints were removed in order to check the stability. After further refinement, deviations were insignificant. The final $R$-factor is significantly better for the---therefore preferred---$C_i$ model (Fig. \ref{fig:1} and Table \ref{tab:1}), in accordance with KY, but at variance with Machida et al. The $R$-factor is slightly improved by further splitting protons in two sub-sites on either sides of dimer planes, separated by $\approx 0.05$ \AA. However, the large overlap of the thermal ellipsoids casts doubt on this splitting, so it is ignored in the remainder of this paper. 

Positional and thermal parameters for heavy atoms in Tables \ref{tab:2} and \ref{tab:3} are quite consistent, within estimated errors, with those of KY, but more accurate. By contrast, the proton positions of KY, $x/a =0.0323(32)$ and $z/c = -0.4102(123)$, are at variance with those in Table \ref{tab:2} and the X-ray based O---H bond length of 0.762(52) \AA\ is too short. These discrepancies could arise from the difference between the maximum of the electronic and nucleus density. On the other hand, O---H bond lengths in Table \ref{tab:4} are virtually identical to those of Machida et al. for the $C_{2h}$ model, and we have checked that the $C_{2h}$ and $C_i$ models refined with our data lead to identical proton positions. Furthermore, proton positions are identical to those previously reported at 14 K and 300 K \cite{FCKeen, FCG2}, although with different occupancies. 

The phase transition can be featured as follows: (i) the discontinuous change of the occupancy of proton sites, with no significant change in positions; (ii) small changes of K and O atom positions by $\approx 0.1$ \AA\ parallel to $(b)$; (iii) the splitting of carbonate groups. The split carbon atoms separated by $\approx 0.2$ \AA\ along ($b$) can be distinguished, while the split oxygen atoms are practically indiscernible. The CO single and double bond lengths (1.395 and 1.210 \AA, respectively in Table \ref{tab:4}) are markedly different from one another and from those at 300 K (1.333 and 1.273 \AA, respectively \cite{FCG2}). 

\section{\label{sec:3} Total neutron scattering: measurements}

In order to observe whether the enhanced features previously attributed to macroscopic entanglement survive above $T_c$, we have conducted diffraction experiments with the SXD instrument at the ISIS pulsed neutron source \cite{SXD,SXD2}. The twinning-free crystal previously measured at 300 K ($\approx 15\times 10 \times 3$ mm$^3$) was loaded in a vacuum tank at $(340 \pm 1)$ K. It was shown that multiple scattering effects are negligible \cite{FCG2}. 

Compared to the four-circle diffractometer, the advantage of SXD is twofold. Firstly, with the high flux of epithermal neutrons at the spallation source, one can measure a more extended domain of reciprocal space than at a reactor source. Secondly, with the time-of-flight technique, the total scattering function is measured at once and signals in addition to Bragg peaks are effectively collected. 

\begin{figure}
\begin{center}
\includegraphics[angle=0.,scale=0.5]{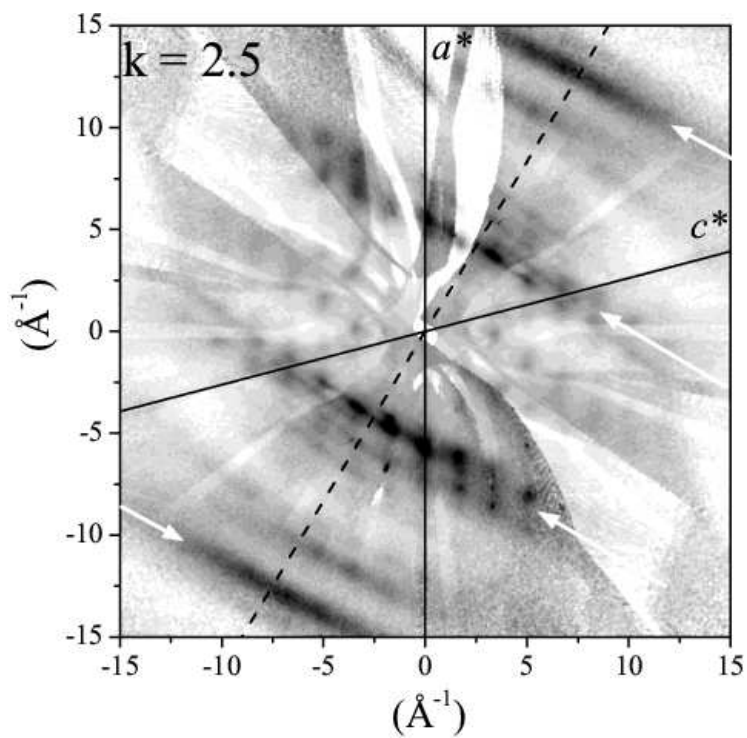}
\end{center}
\caption{\label{fig:3} Map of scattering intensity for KHCO$_3$ at 340 K in between  ($a^*, c^*$) reciprocal planes. The arrows emphasize ridges of intensity parallel to $(z)$ and perpendicular to dimer planes.}
\end{figure}

By definition, $Q_x$, $Q_y$, $Q_z$, are projections of $\mathbf{Q}$ onto the normal mode axes for protons, $x$, $y$, $z$ (Fig. \ref{fig:1}). The maps of intensity (Figs \ref{fig:2} and \ref{fig:3}) for $(a^*,c^*)$ planes at different $k$-values ($k = Q_y/b^*$, with $b^* \approx 1.112$ \A) are concatenations of measurements for 8 different crystal orientations, with $(b) \parallel (b^*)$ either parallel or perpendicular to the equatorial plane of the detector bank. 

The anisotropic incoherent scattering intensity centered at $\mathbf{Q} = 0$, that was clearly observed at 15 K \cite{FCKeen}, is barely visible at 340 K. The shortfall of intensity with $|\mathbf{Q}|$ is a consequence, via the Debye-Waller factor, of the thermal population of the density-of-states \cite{FCG2,CCW}. 

Cigar-like shaped ridges of intensity along $Q_z$ are visible in Fig. \ref{fig:2} for particular $k$-values. For $k = 0$, they appear at $Q_x = 0$ and $\pm (10.00 \pm 0.25)$ \A. For $k = 1$, they are barely visible. For $k = 2$ or 3, we observe a different pattern of ridges at $Q_x = \pm (5 \pm 0.2)$ and $\pm (15 \pm 0.2)$ \A, still along $Q_z$. These features are best visible in Fig. \ref{fig:3} for $k = 2.5$. There is no visible ridge at $k = 4$. Then, from $k = 5$ to 9, we observe the same sequence as for $k = 0$ to 4. These ridges are basically identical to those previously reported at 300 K or below for the ordered $P2_1/a$ crystal \cite{FCKeen,FCG2}. Consequently, they are not due to transition induced phenomena, such as disorder, or domains, or planar faults, or critical behaviour,  for which coherent diffuse scattering should vanish below $T_c$ \cite{NK}. Below it is shown that these features can be attributed to diffraction by the sublattice of entangled protons. 

\section{\label{sec:4} Theory}

\subsection{\label{sec:41}The adiabatic separation}

Within the framework of the Born-Oppenheimer approximation, the vibrational Hamiltonian can be partitioned as
\begin{equation}\label{eq:1}
\mathcal{H}_\mathrm{v} = \mathcal{H}_{\mathrm{H}} +\mathcal{H}_{\mathrm{at}}+ \mathcal{C}_{\mathrm{Hat}},
\end{equation}
where $\mathcal{H}_{\mathrm{H}}$ and $\mathcal{H}_{\mathrm{at}}$ represent the sublattices of protons (H$^+$) and heavy atoms, respectively, while $\mathcal{C}_{\mathrm{Hat}}$ couples the subsystems. For OH$\cdots$O hydrogen bonds, coupling terms between OH and O$\cdots$O degrees of freedom are rather large \cite{Novak,SZS}, and beyond the framework of the perturbation theory. Two approaches, either semiclassical or quantum, are commonly envisaged.

Semiclassical protons are dimensionless particles, with definite positions and momentums, moving across a potential hypersurface \cite{BVIT,BVT,GPNK,SFS2,TVL}. Complex trajectories involving heavy atoms lead to mass renormalization, and to incoherent phonon-assisted tunnelling \cite{ST,EGS,BHT}. This approach is quite natural when the Born-Oppenheimer surface is calculated from first principles, but quantum effects are severely underestimated. 

Alternatively, if the classical concept of ``trajectory'' is abandoned, adiabatic separation of $\mathcal{H}_{\mathrm{H}}$ and $\mathcal{H}_{\mathrm{at}}$ may lead to tractable models \cite{FCG2,SZS,GPNK,FRLL,MW,W}. Then, protons in a definite eigen state should remain in the same state in the course of time, while heavy atoms oscillate slowly, in an adiabatic hyperpotential depending on the proton state, through the coupling term. This separation is relevant for KHCO$_3$ because adiabatic potentials for different protons states do not intersect each other. In fact, the separation is rigorously exact in the ground state, since protons should remain in this state for ever, if there is no external perturbation. Then, protons are bare fermions and quantum correlations arise \cite{FCG2}. 

\subsection{\label{sec:42}Macroscopic proton states}

Consider a crystal composed of very large numbers $N_a$, $N_b$, $N_c$ ($\mathcal{N}=N_{a}N_{b}N_{c}$) of unit cells labelled $j,k,l,$ along crystal axes $(a),$ $(b),$ $(c)$, respectively. Because centrosymmetric dimers have no permanent electric dipole, interdimer coupling terms are negligible \cite{FTP,IKSYBF,KIN}. Proton dynamics for a $C_i$ dimer is modelled with collinear oscillators in three dimensions, along coordinates $\alpha_{1{jkl}}$ and $\alpha_{2{jkl}}$ ($\alpha =x,y,z$) with respect to equilibrium positions at $\pm\alpha_{0{jkl}}$. The mass-conserving normal coordinates and their conjugated momentums, 
\begin{equation}\label{eq:2}
  \begin{array}{lc}
\alpha_{{s}} = \displaystyle{\frac {1} {\sqrt{2}}\left(\alpha_{1} - \alpha_{2} + 2\alpha_0 \right)}, & P_{{s}\alpha} = \displaystyle{\frac {1} {\sqrt{2}} \left( P_{1\alpha} - P_{2\alpha} \right)}, \\
    \alpha_{{a}} = \displaystyle{\frac {1} {\sqrt{2}}\left(\alpha_{1} + \alpha_{2} \right)}, & P_{\mathrm{a}\alpha} = \displaystyle{\frac {1} {\sqrt{2}}\left( P_{1\alpha} + P_{2\alpha } \right)},
  \end{array}
\end{equation}
lead to uncoupled oscillators at frequencies $\hbar\omega_{\mathrm{s}\alpha}$ and $\hbar\omega_{{a}\alpha}$, respectively, each with $m = 1$ amu. The difference $\hbar\omega_{{s}\alpha} - \hbar\omega_{{a}\alpha}$ depends on the coupling term (say $\lambda_\alpha$). The wave functions, namely $\Psi_{{njkl}}^{a}( \alpha_{{a}})/\sqrt{2}$, $\Psi_{{n'jkl}}^{s}(\alpha_{{s}} -\sqrt{2}\alpha_{0})/\sqrt{2}$ for a half-dimer, cannot be factored into wave functions for individual particles, so there is no local information available for these entangled oscillators and the degenerate ground state must be antisymmetrized. For this purpose, the wave functions are rewritten as linear combinations of those for permuted oscillators as
\begin{equation}\label{eq:3}
\displaystyle{\frac{1} {\sqrt{2}}} \Theta_{0{jkl}\pm } = \displaystyle{\frac{1} {2}} \prod\limits_\alpha \Psi_{0{jkl}}^{a} (\alpha_{{a}}) \left [ \Psi_{0{jkl}}^{s}(\alpha_{{s}} -\sqrt{2}\alpha_{0}) \pm \Psi_{0{jkl}}^{s} (\alpha_{{s}} +\sqrt{2}\alpha_{0}) \right ],
\end{equation}
where $\Theta_{0{jkl}\pm }$ is the total wave function for superposed half-dimers with identical proton sites. Then, the antisymmetrized state vectors are: 
\begin{equation}\label{eq:4}
\begin{array}{rcl}
|0jkl+ \rangle \otimes |S\rangle & = & \big | \Theta_{0{jkl}+} \rangle \otimes \displaystyle{\frac{1}{\sqrt{2}}} \left [ |\uparrow_1 \downarrow_2 \rangle - | \downarrow_1 \uparrow_2 \rangle \right ] ;\\
|0jkl- \rangle \otimes |T\rangle & = & | \Theta_{0{jkl}-} \rangle \\
& \otimes & \displaystyle{\frac{1}{\sqrt{3}}} \left [ | \uparrow_1 \uparrow_2 \rangle + | \downarrow_1 \downarrow_2 \rangle + \displaystyle{\frac{1}{\sqrt{2}}} [ |\uparrow_1 \downarrow_2 \rangle + |\downarrow_1 \uparrow_2 \rangle ] \right ] .
\end{array} 
\end{equation}
The oscillators are now entangled in position, momentum, and spin. In contrast to magnetic systems \cite{Cowley}, there is no level splitting, so the symmetry-related entanglement is energy-free. It is also independent of the magnitude of the coupling terms between oscillators. Furthermore, in contrast to Keen and Lovesey \cite{KL1}, or Sugimoto et al. \cite{SOY}, we argue, as an experimental fact, that there is no significant exchange integral for protons separated by $\approx 2.2$ \AA\ \cite{FCou}. Protons in KHCO$_3$ are definitely not itinerant particles with an energy band structure. 

The spatial periodicity of the crystal leads to extended states and nonlocal observables in three dimensions. Consider the vibrational wave function for a unit cell containing two dimer entities as  $\Xi_{0{jkl}\tau} = \Theta_{0{jkl}\tau} \pm \Theta'_{0{jkl}\tau}$, with $\tau =$ ``$+$'' or ``$-$''. Then, trial phonon waves can be written as 
\begin{equation}\label{eq:5}
\Xi_{0\tau} (\mathbf{k})= \displaystyle{\frac{1}{\sqrt {\mathcal{N}}}}  \sum\limits_{{l} = 1}^{{N_c}} \sum\limits_{{k} = 1}^{{N_b}} \sum\limits_{{j} = 1}^{{N_a}} \Xi_{0{jkl}\tau} \exp(i\mathbf{k\cdot L}), 
\end{equation}
where $\mathbf{k}$ is the wave vector and $\mathbf{L}  = j \mathbf{a} + k \mathbf{b} + l \mathbf{c}$, with the unit cell vectors $\mathbf{a}$, $\mathbf{b}$, $\mathbf{c}$. For fermions, antisymmetrization leads to 
\begin{equation}\label{eq:6}
\mathbf{k\cdot L} \equiv 0 \mathrm{\ modulo\ } 2\pi. 
\end{equation}
Consequently, there is no phonon (no elastic distortion) in the ground state. This symmetry-related ``super-rigidity'' \cite{FCG2} is independent of proton--proton interaction. Then, state vectors in three dimensions can be written as: 
\begin{equation}\label{eq:7}
\begin{array}{c}
\left | \Xi_{0 +} (\mathbf{k = 0}) \right \rangle \otimes |S \rangle; \\
\left | \Xi_{0 -} (\mathbf{k = 0}) \right \rangle \otimes |T \rangle. 
\end{array}
\end{equation}
There is no local information available for these entangled states and the wave functions $\Xi_{0 \tau} (\mathbf{k = 0})$ describes oscillations of the super-rigid sublattice as a whole. Finally, the ground state of the sublattice is:
\begin{equation}\label{eq:8}
\begin{array}{c}
\sqrt{\mathcal{N}} | \Xi_{0 +} (\mathbf{k = 0}) \rangle \otimes |S \rangle;\\ 
\sqrt{\mathcal{N}} | \Xi_{0 -} (\mathbf{k = 0}) \rangle \otimes |T \rangle. 
\end{array}
\end{equation}
This state is intrinsically steady against decoherence. Irradiation by plane waves (photons or neutrons) may single out some excited states. Entanglement in position and momentum is preserved, while the spin-symmetry and super-rigidity are destroyed. However, these features are reset automatically after decay to the ground state, presumably on the time-scale of proton dynamics. This mechanism allows the super-rigid sublattice to be at thermal equilibrium with the surroundings, despite the lack of internal dynamics. Note that even at room temperature, the thermal population of the first excited proton state ($\gamma$OH $\approx 1000$ \cm) is negligible ($< 1\%$). 

\section{\label{sec:5} Differential cross section: calculations}

\begin{figure}[!hbtp]
\begin{center}
\includegraphics[scale=0.25]{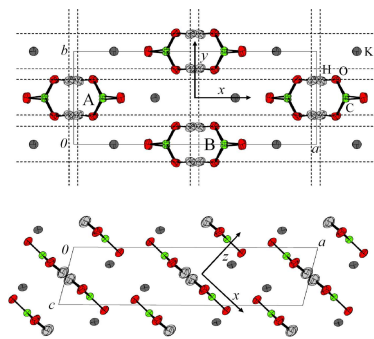}
\end{center}
\caption{\label{fig:1} Schematic view of the crystalline structure of KHCO$_{3}$ at 340 K. The thin solid lines represent the unit cell. Top: projection onto the plane normal to $c$. The dotted lines joining protons emphasize the proton sublattice (see text). Bottom: projection onto $(a,c)$. }
\end{figure}

\begin{figure}
\includegraphics[angle=0.,scale=0.3]{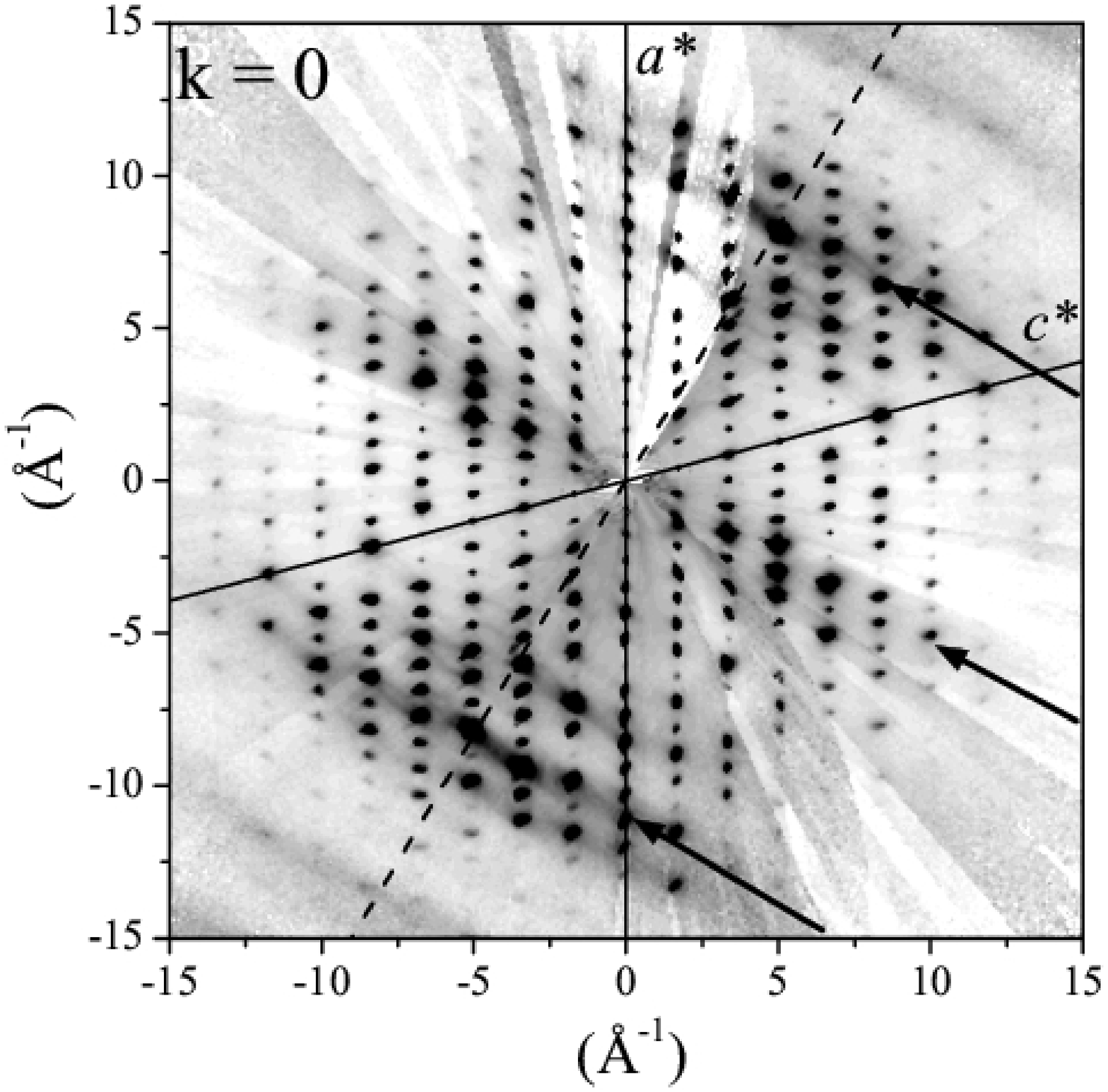}
\includegraphics[angle=0.,scale=0.3]{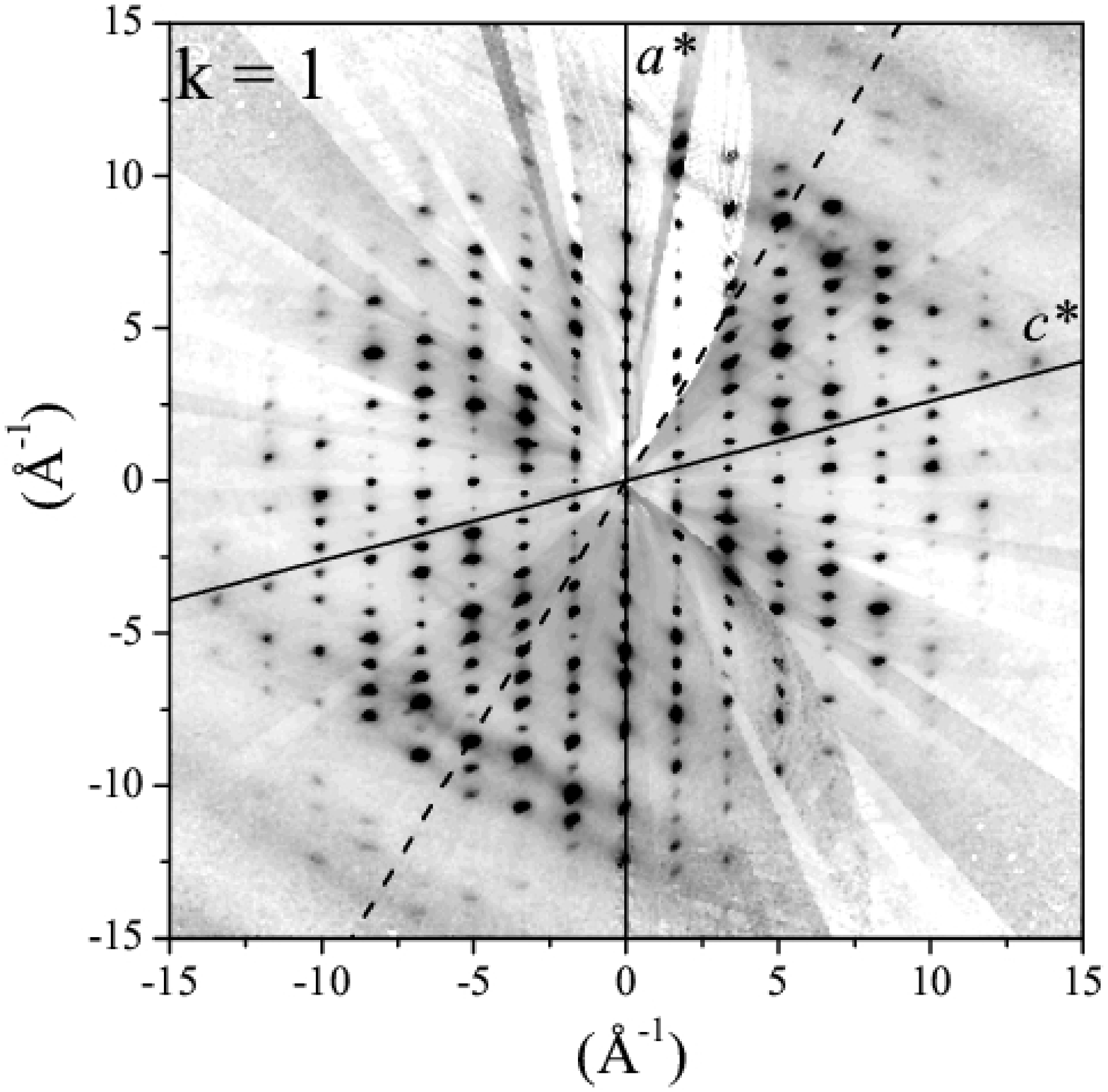}
\includegraphics[angle=0.,scale=0.3]{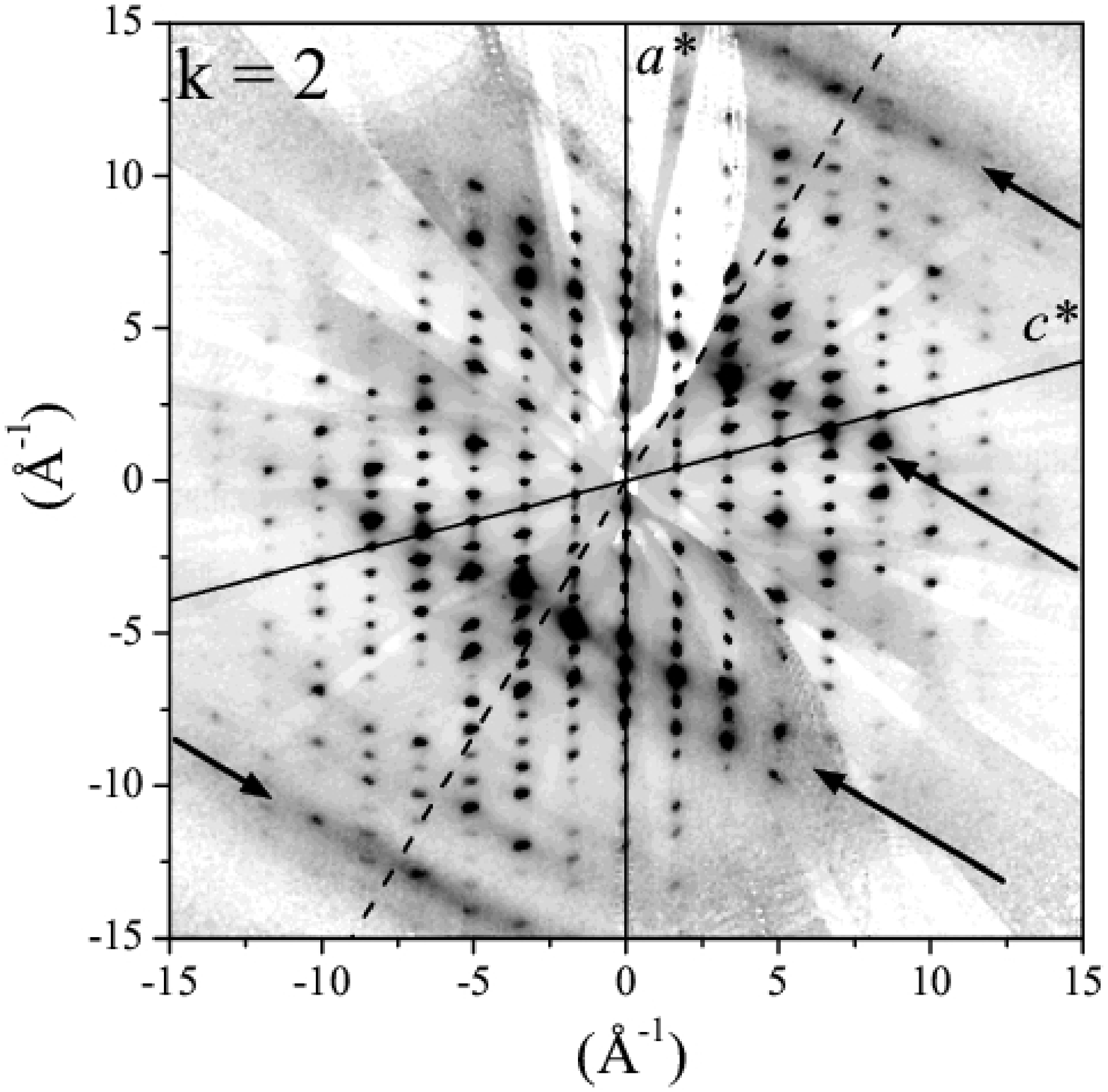}
\includegraphics[angle=0.,scale=0.3]{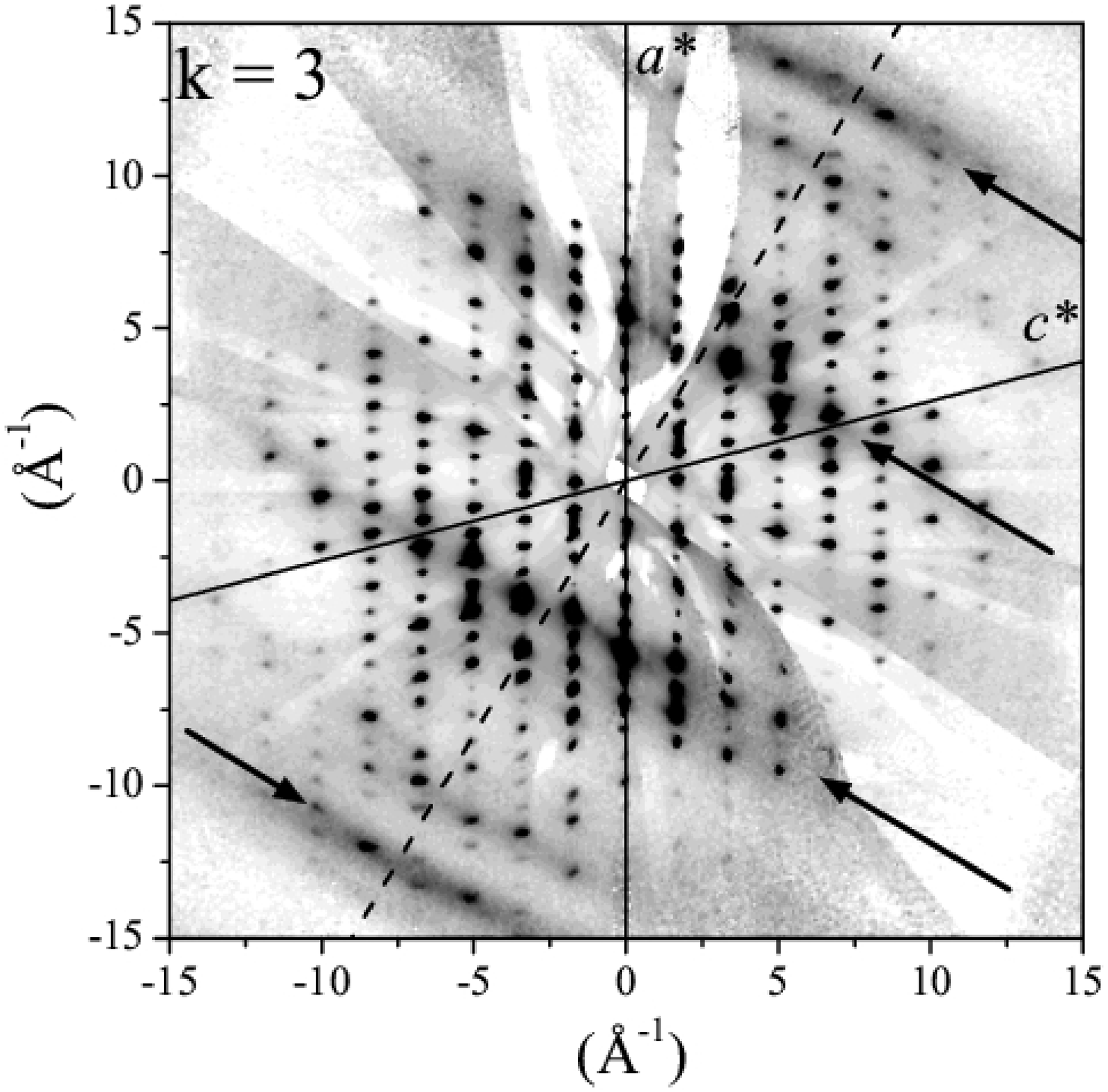}
\includegraphics[angle=0.,scale=0.3]{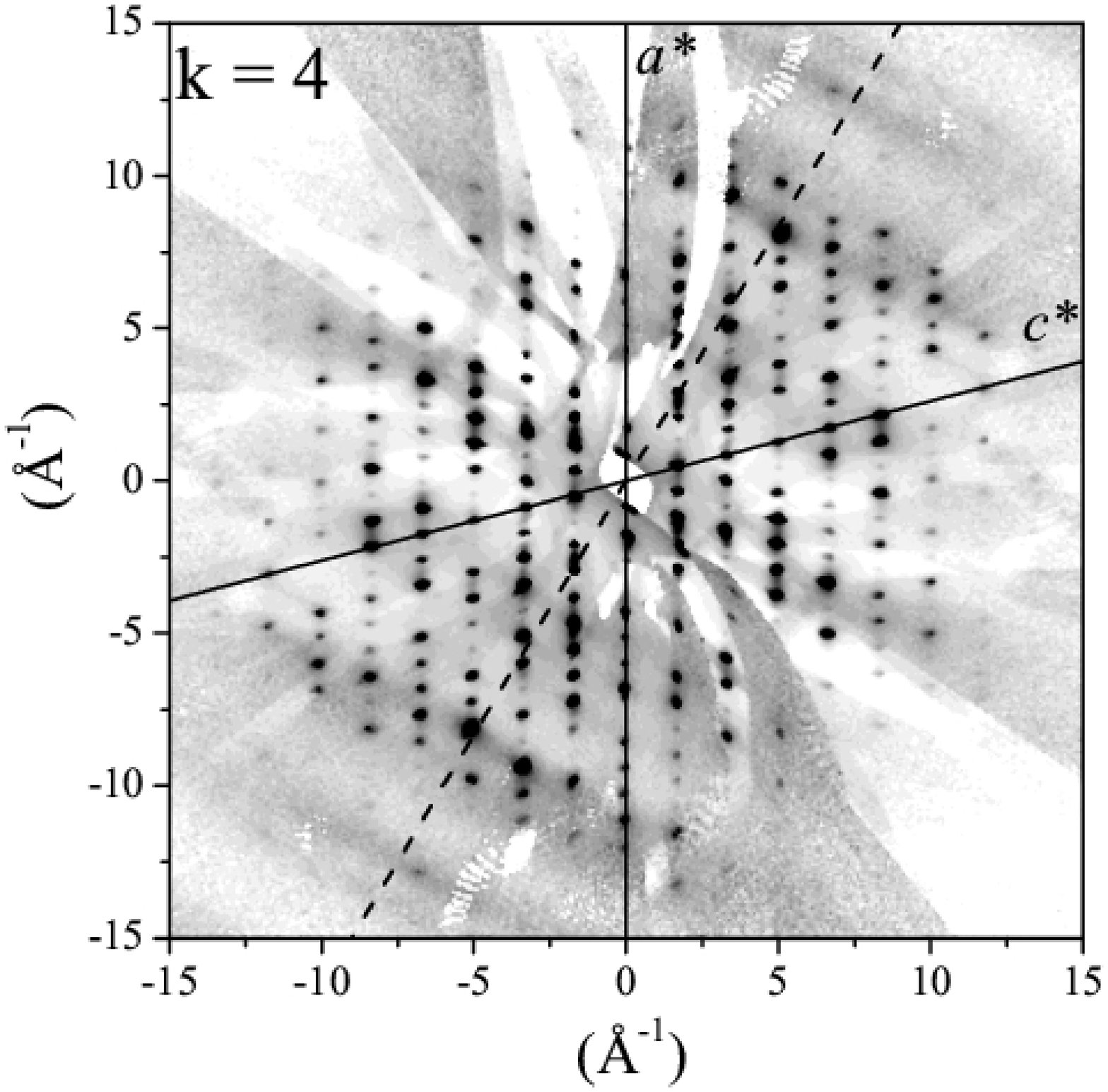}
\includegraphics[angle=0.,scale=0.3]{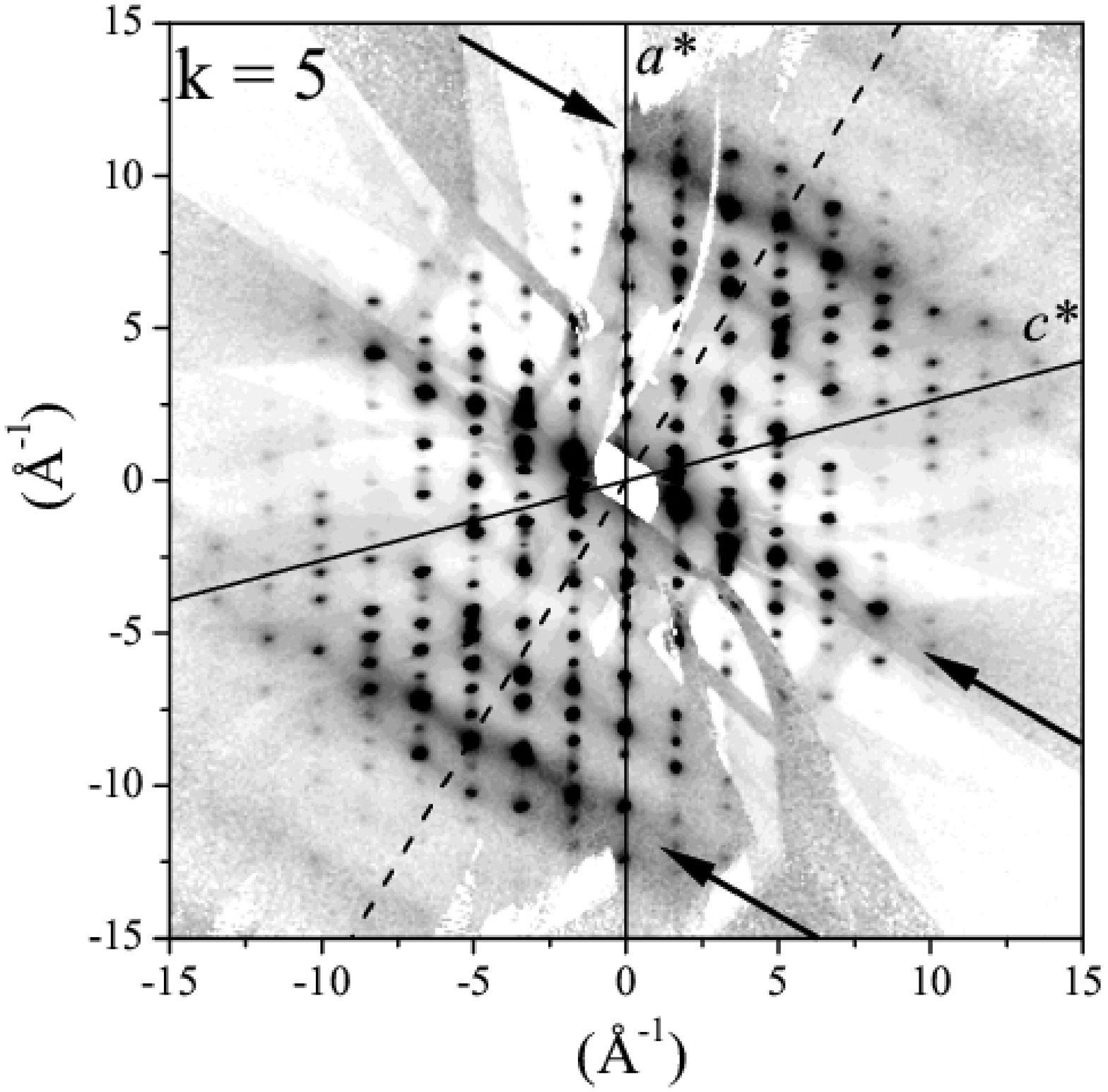}
\includegraphics[angle=0.,scale=0.3]{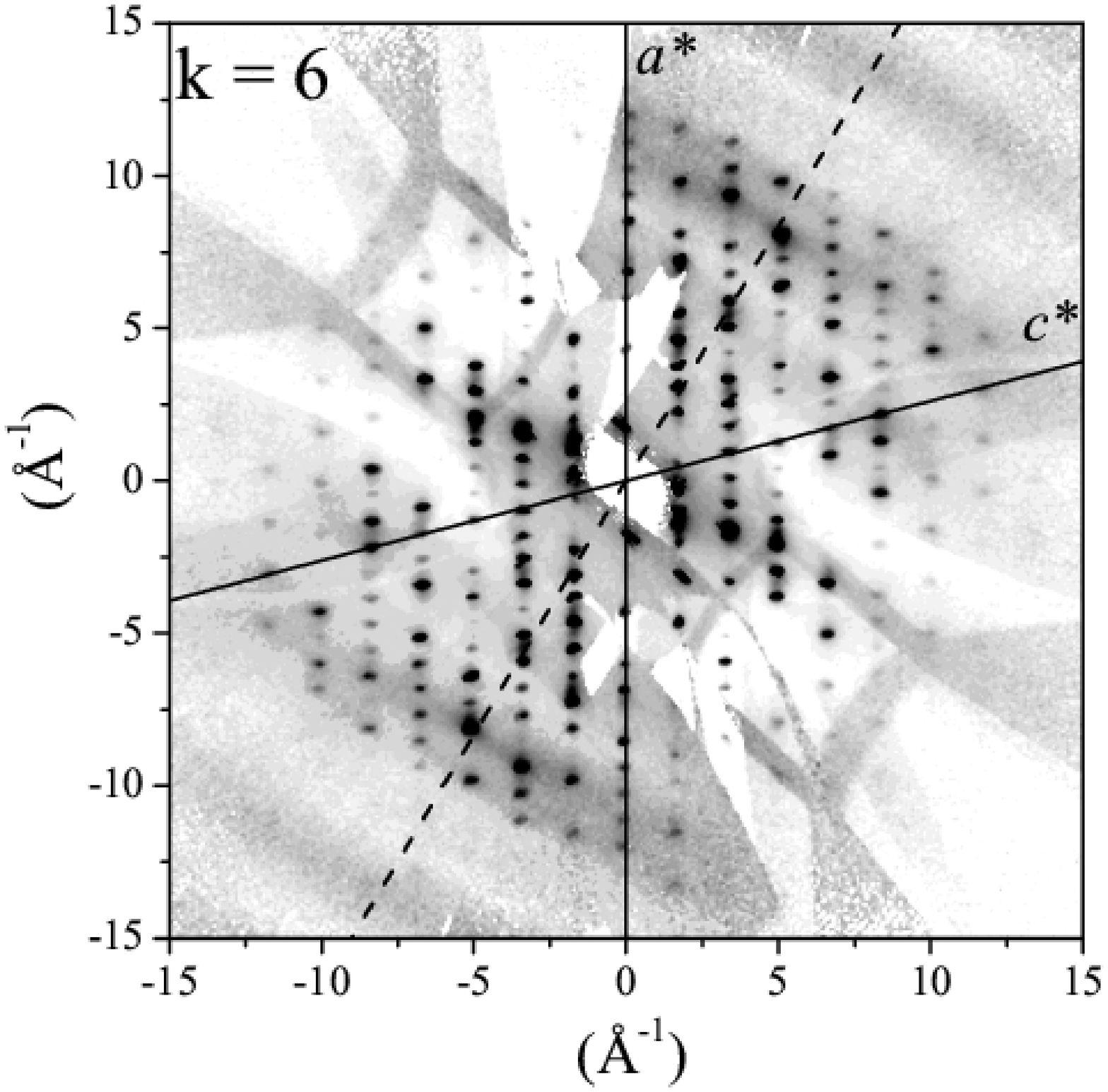}
\includegraphics[angle=0.,scale=0.3]{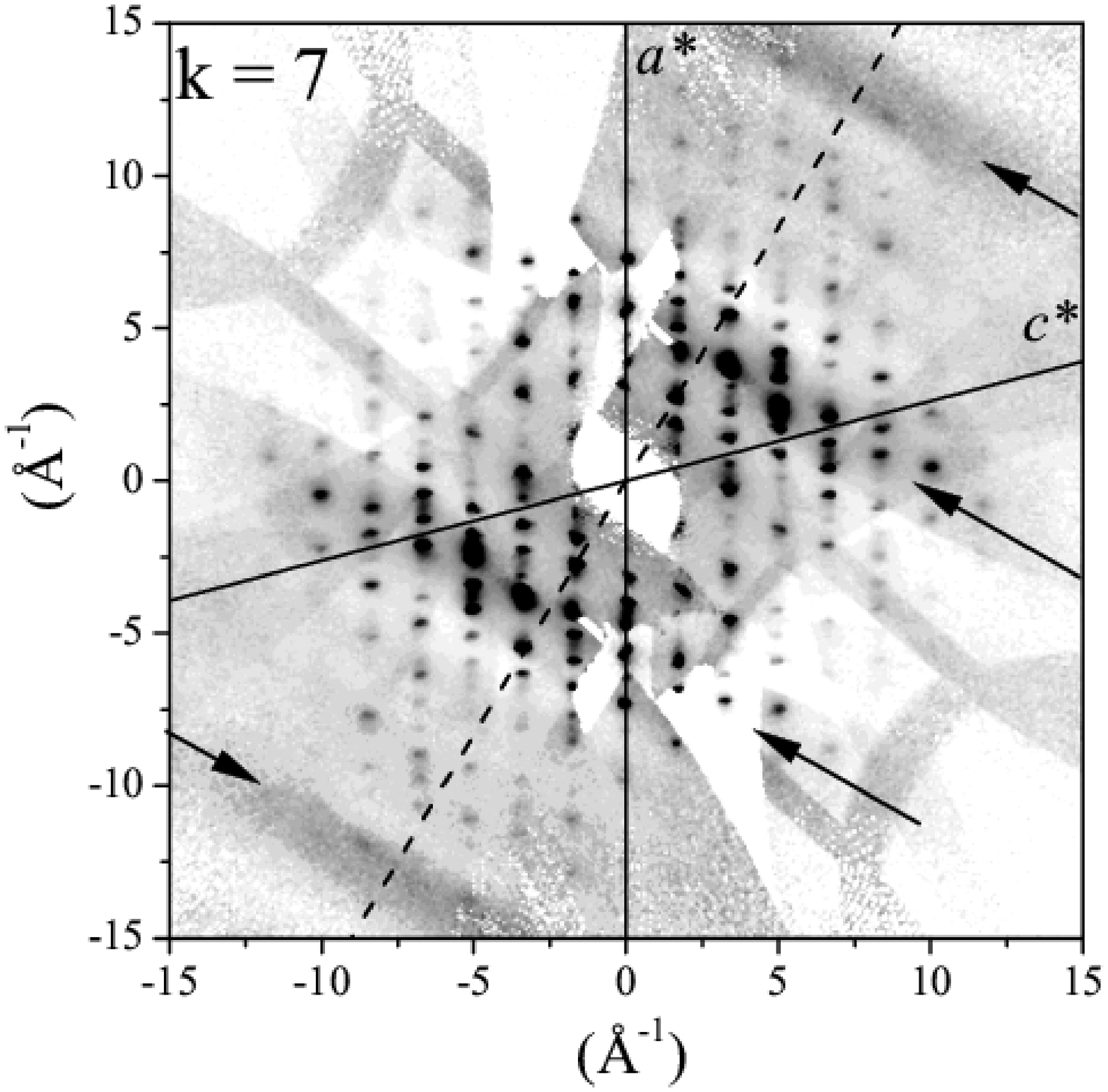}
\includegraphics[angle=0.,scale=0.3]{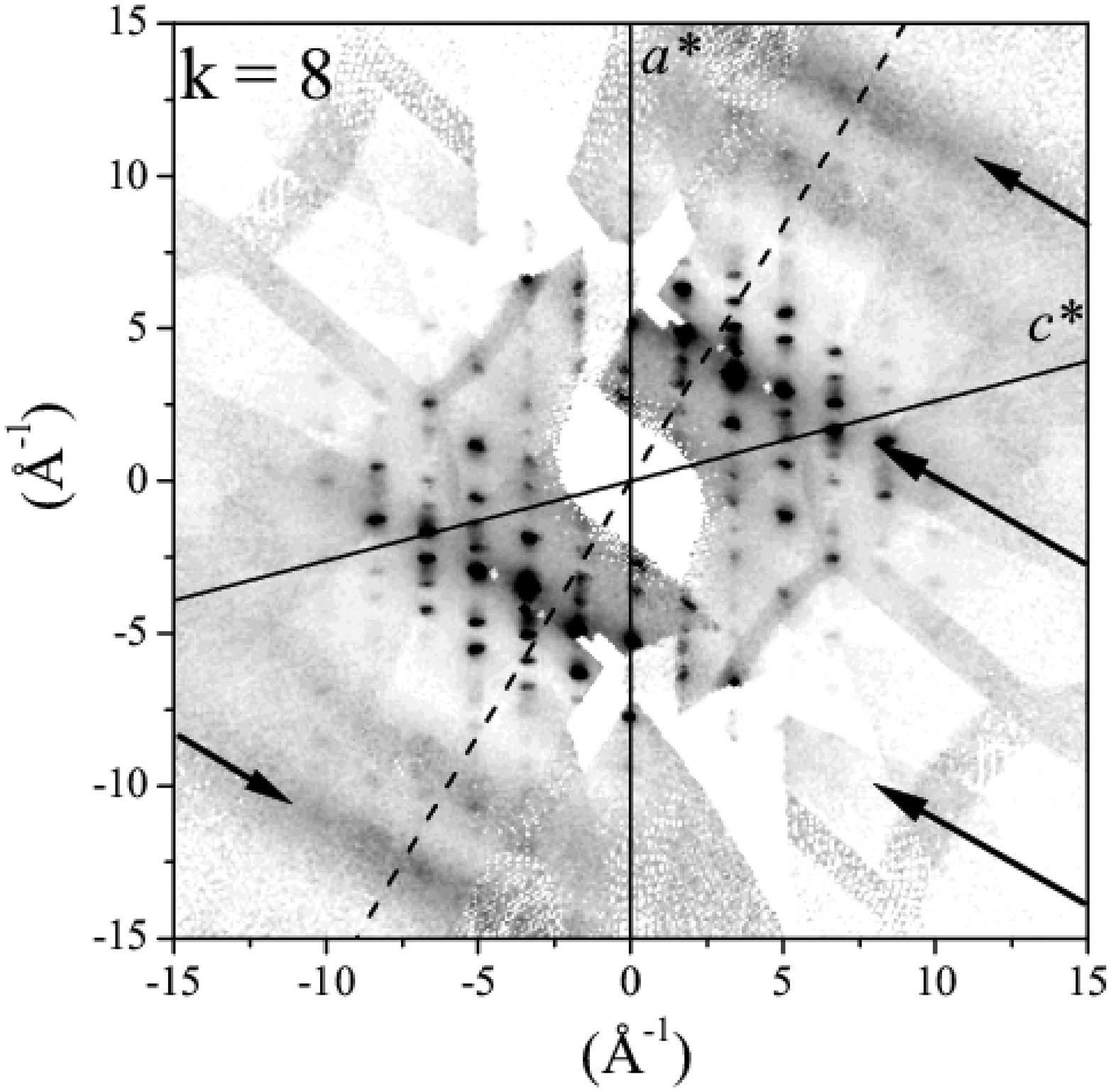}
\caption{\label{fig:2} Diffraction patterns of KHCO$_3$ at 340 K in ($a^*, c^*$) planes. The arrows point to the ridges of intensity for momentum transfer parallel to $(z)$ and perpendicular to the trace of dimer planes (dash lines), which contain the $(x)$ direction, as defined in Fig. \ref{fig:1}. The $(y)$ direction parallel to ($b^*$) is perpendicular to $(a^*,c^*)$. $k = Q_y/b^*$ (see Table \ref{tab:5}).}
\end{figure}

Neutrons (spin $1/2$) are unique to observing the spin-symmetry of proton states (\ref{eq:7}). However, the spin-symmetry is not stabilized by any energy. It is therefore extremely fragile and only ``noninvasive'' experiments, free of measurement-induced decoherence, can evidence quantum entanglement \cite{LG}. For neutron scattering, this means (i) no energy transfer (ii) no spin-flip and (iii) particular $\mathbf{Q}$-values preserving the super-rigidity. For planar dimers, these conditions are realized when $Q_x$, $Q_y$, match a node of the reciprocal sublattice of protons, so that scattering events probe the super-rigid lattice without any perturbation of its internal state. The spin-symmetry is probed along the neutron-spin direction without spin-flip and the cross-section, namely the total nuclear cross-section $\sigma_H$, is $\approx 45$ times greater than the coherent cross-section, $\sigma_{Hc}$, when the matching conditions are not realized \cite{FCG2,SWL}. Enhanced diffraction probe the perfect periodicity of the super-rigid sublattice, so the Debye-Waller factor is equal to unity. Alternatively, when the matching conditions are not realized, phonons created by neutrons contribute to the thermal parameters in Table \ref{tab:3} and to the ellipsoids in Figs \ref{fig:1}. 

In fact, enhanced diffraction and Bragg peaks of heavy atoms are superimposed. Their intensities are proportional to $\sigma_H$ and $\sigma_{cKCO_3} \exp\{-2W_{KCO_3}(\mathbf{Q})\}$, respectively, where $\sigma_{cKCO_3} \approx 27.7$ bn is multiplied by the Debye-Waller factor. At large $\mathbf{Q}$-values, the contribution of heavy atoms is largely hidden by that of correlated protons, and this effect is more pronounced at elevated temperatures. 

The dashed lines in Fig. \ref{fig:1} show that proton sites are aligned along $x$ and $y$, but not along $z$. Consequently, the noninvasive condition can be realized for $Q_\mathrm{x}$ and $\ Q_\mathrm{y}$ exclusively, whereas $Q_\mathrm{z}$ never coincides with a nod of the reciprocal lattice of protons. Then, the diffraction pattern depends on whether we consider incoherent or coherent scattering along $Q_\mathrm{z}$. 

In order to calculate the differential scattering functions, we consider a $C2/m$ super-rigid sublattice composed of primitive unit cells $(a/2,b,c)$ with $(j,k,l)$ indexes. Proton sites are at $\pm \{x_{0jkl}, y_{0jkl}, 0\}$, $\pm \{x_{0jkl}, -y_{0jkl}, 0\}$, with respect to the centre of split dimers. The periodicity of the grating-like structure parallel to $y$ is $D_\mathrm{x}/2$, with $D_\mathrm{x} \approx a/\cos 42 ^\circ \approx 20.39$ \AA, while the periodicity of double-lines parallel to $x$ is $D_\mathrm{y} = b$. 

\subsection{\label{sec:51} Super-rigid arrays in two dimensions}

The differential cross-section can be written as 

\begin{equation}\label{eq:9}
\begin{array}{rcl}
\displaystyle{\frac{d\sigma_{2}}{d\Omega}} & \propto & \sum\limits_{l =1} ^{N_c} \sum\limits_{\tau_{i}} \sum\limits_{\tau_{f}} \left | \sum\limits_{j = 1} ^{N'_a} \sum\limits_{k =1} ^{N_b} \left\{ \left [ \exp iQ_y \left(kD_y - y_0 \right) \right. \right. \right. \\
& + & \left. \tau_{i} \tau_{f} \exp iQ_y \left(kD_y + y_0 \right ) \right ] \\
 & \times & \left. \left [ \exp i Q_x \left(jD_x/2 - x_0 \right ) + \tau_{i}\tau_{f} \exp i Q_x \left(jD_x/2 + x_0 \right ) \right ] \right\}^2 \Bigg |^2 \\
& \times & \exp-2W_z (Q_z), 
\end{array} 
\end{equation}
with $N'_a = 2N_a$. Neutrons are scattered either in-phase ($\tau_f \tau_i = +1$) or anti-phase $(\tau_f \tau_i = -1)$ by pairs of lines separated by $2x_0 \approx 0.6$ \AA\ and $2y_0 \approx 2.209$ \AA, respectively. The phase matching condition, namely $x_0$ ($y_0$) commensurable with $D_x/2$ ($D_y$), is intrinsic to the crystal structure. $\{\cdots\}^2$ accounts for simultaneous scattering by superposed singlet-like and triplet-like states (\ref{eq:7}), without neutron-spin flip. The compound Debye-Waller factor $\exp-2W_\mathrm{z} (Q_\mathrm{z})$, including contributions from all atoms, accounts for incoherent scattering along $Q_\mathrm{z}$. 

The diffraction pattern is composed of rods of diffuse scattering parallel to $Q_z$, cigar-like shaped by the Debye-Waller factor, at $Q_x$, $Q_y$-values corresponding to divergences of (\ref{eq:9}) gathered in table \ref{tab:5}. Such divergences occur at $Q_y = n_y \pi /y_0 \approx n_y \times 2.86$ \A, since $Q_y D_y/\pi \approx 5 n_y$ is integer. There is no divergence for anti-phase scattering at $Q_y = \pm (n_y + 1/2) \pi/y_0$, because $Q_yD_y/\pi \approx 5 (n_y + 1/2)$ is not integer. For $n_y$ even, $Q_yD_y/\pi$ is also even, $\tau_{i} = \tau_{f}$, and ridges are anticipated at $Q_x = n_x \pi/x_0 \approx n_x\times 10$ \A, since $Q_x D_x/\pi \approx 68n_x$ is even. Alternatively, for $n_y$ odd, $Q_yD_y / \pi$ is also odd, $\tau_{i} \neq \tau_{f}$, and ridges are anticipated at $Q_x = (n_x + 1/2) \pi/x_0$, since $Q_x D_x/\pi \approx 68(n_x + 1/2) $ is even. In this case, the enhanced features are best visible at $k \approx 2.57$ (Fig. \ref{fig:3}). 

\subsection{\label{sec:52} Super-rigid lattice in three dimensions}

The differential cross-section for no spin-flip scattering events is 
\begin{equation}\label{eq:10}
\begin{array}{rcl}
\displaystyle{\frac{d\sigma_{3}}{d\Omega}} & \propto & \sum\limits_{\tau_{i}} \sum\limits_{\tau_{f}} \left | \sum\limits_{j = 1} ^{N'_a} \sum\limits_{k =1} ^{N_b} \sum\limits_{l =1} ^{N_c} \exp i Q_z l D_z \right. \left\{\left [ \exp iQ_y \left(kD_y - y_0 \right)\right.\right.\\
& + & \left. \tau_{i}\tau_{f} \exp iQ_y \left(kD_y + y_0 \right ) \right ] \\
& \times & \left. \left [\exp i Q_x \left(jD_x/2 - x_0 \right ) + \tau_{i}\tau_{f} \exp i Q_x \left(jD_x/2 + x_0 \right ) \right ] \right\}^2 \Bigg |^2 . 
\end{array} 
\end{equation}
Divergences occur along the previous rods of intensity at $Q_z = \pm n_z 2\pi / D_z$, with $D_z \approx c\times \cos 28^\circ \approx 3.28$ \AA\ and $2 \pi /D_z \approx 1.92$ \A. In Fig. \ref{fig:2}, these peaks clearly emerge from the rods of diffuse scattering depressed by the Debye-Waller factor in (\ref{eq:9}). 

Equations (\ref{eq:9}) and (\ref{eq:10}) confirm that the enhanced features correspond to the reciprocal sublattice of protons represented by state vectors (\ref{eq:8}) based on a theoretical framework free of ad hoc hypothesis or parameters. The only prerequisite, namely the adiabatic separation, is an intrinsic property of hydrogen bonds in this crystal. By contrast, other models \cite{KL1,SOY} cannot account for enhanced diffraction. 

\section{\label{sec:6}Discussion}

Kashida and Yamamoto \cite{KY} have opposed the ordered $C_{2h}$ to the disordered $C_i$ structures. The argument consists in that for the former there is no splitting of heavy atom sites and the symmetrized double-wells for protons lead to quantum delocalization \cite{EGS,KSNZKY,LJ,LJ2,Odin}. By contrast, split $C_i$ dimers are supposed to be representative of a statistical distribution of atom positions and the double well remains asymmetric, as it is below $T_c$. 

Kashida and Yamamoto \cite{KY} or Eckold et al. \cite{EGS} have observed that the $P2_1/a$ superlattice reflections ($h+k = 2n+1$) merge above $T_c$ into elastic diffuse scattering, cigar-like shaped along ($a^*$). The intensity decreases as the temperature increases but the signal is still visible above $340$ K. According to KY, the $P2_1/a$ local structure could survive above $T_c$ as pre-transitional domains whose correlation length diverges as $T$ goes down to $T_c$. Diffuse scattering along $(a^*)$ is invisible with SXD (Fig. \ref{fig:2}) but this does not question the existence of this signal. 

Critical phenomena have been also thoroughly investigated for the phase transition of KDCO$_3$ at $T_{cD} = 353$ K, taking advantage of the larger coherent scattering cross-section of the deuterium atom \cite{KNHKKY1,KNHKKY2,KSNZKY}. QENS measurements at the superlattice reflection $(1,2,0)$ with the high resolution spin-echo technique evidence very slow dynamics ($\sim 10^{-9} s$) above $T_{cD}$ and complete slowing down as $T \longrightarrow T_{cD}$. These dynamics were tentatively associated with pre-transitional collective reorientation of dimers, eventually correlated with the softening of an acoustic mode propagating along $(a^*)$, with polarization parallel to $(b)$ \cite{H}. On the other hand, diffuse scattering around the main reflection $(0,2,0)$ was attributed to the relaxation of individual dimers \cite{KNHKKY1,KNHKKY2,KSNZKY}. 

In contrast to the order-disorder schemes envisaged in previous works, the diffraction patterns presented in Figs \ref{fig:2} and \ref{fig:3} lead to the conclusion that the structural transition is of the order-order type. Then the question arises as to whether the double-wells undergoes symmetrization or not. 

\begin{figure}
\begin{center}
\includegraphics[angle=0.,scale=0.7]{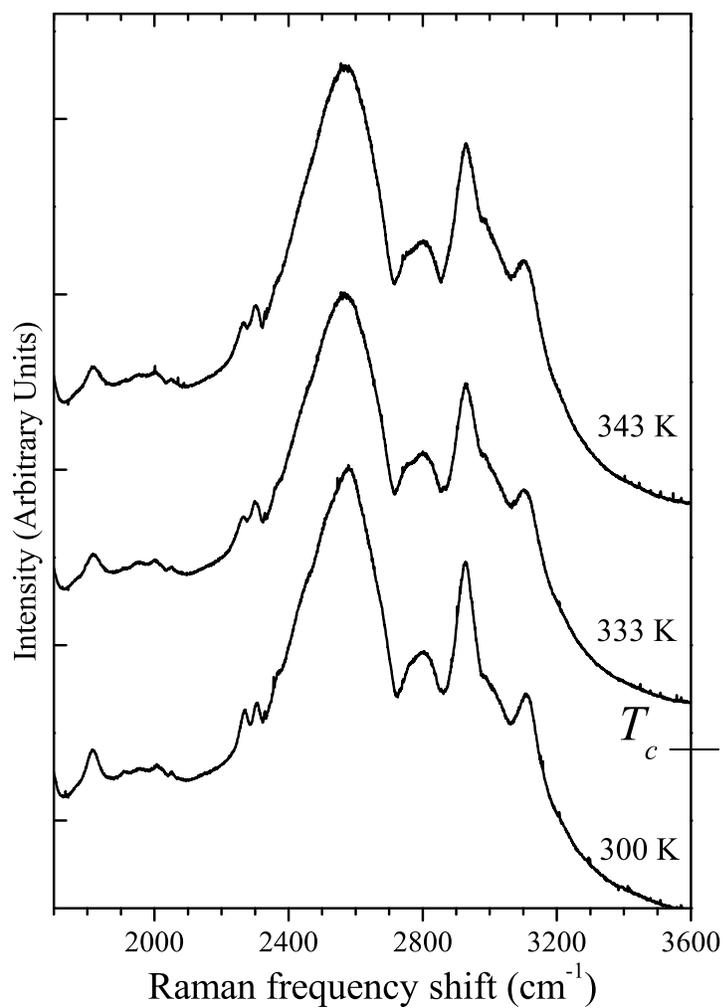}
\caption{\label{fig:5} Raman spectra of powdered KHCO$_3$ crystal across the phase transition at $T_c = 318$ K. The spectra were recorded with a DILOR XY triple-monochromator equipped with an Ar$^+$ laser emitting at 4880 \AA. The spectral resolution was $\approx 2$ \cm. The sample was sealed in a glass tube and then loaded in a furnace. The temperature was controlled within $\pm 2$ K. }
\end{center}
\end{figure}

The shape of the double-wells along the hydrogen bonds can be probed with Raman spectroscopy in the frequency range of the OH stretching vibrations \cite{Fil7,Fil2}. In Fig. \ref{fig:5}, the submaxima observed below $T_c$ at $\approx 2580$, $2800$, $2930$, $3100$ \cm\ are representative of the four transitions for the asymmetric double-wells of the $P2_1/a$ structure. Additional transitions (for example overtones and combinations) cannot be excluded, but a detailed assignment scheme is not necessary because the chief point of the argument is that symmetrization of the double-wells above $T_c$ should be of dramatic consequences to the spectral profile: (i) large frequency shifts due to the collapse of the potential asymmetry; (ii) the symmetry-related selection rules should extinguish two among the four transitions. Figure \ref{fig:5} shows that the extended eigenstates of the $\nu$OH mode probed at the Brillouin-zone-centre are identical for the $P2_1/a$ and $C2/m$ structures. Symmetrization is excluded. 

The $C2/m$ structure can be featured as follows. Firstly, enhanced diffraction suggests that the crystal is in a pure state ($|C2/m\rangle$) composed of extended states for all dynamical degrees of freedom. Secondly, Raman spectra suggest that $|C2/m\rangle$ can be represented with a superposition of state vectors $|P_u\rangle$ and $|P_d\rangle$ representing $P2_1/a$-like substructures, symmetric with respect to $(a,c)$. They are labelled with ``up'' ($u$) and ``down'' ($d$) orientations of a pseudospin along $z$. According to the quantum theory of measurements \cite{Leggett,CTDL,Leggett4}, the eigenstates of either $|P_u\rangle$ or $|P_d\rangle$ are probed with Raman. 

\begin{figure}
\begin{center}
\includegraphics[angle=0.,scale=0.17]{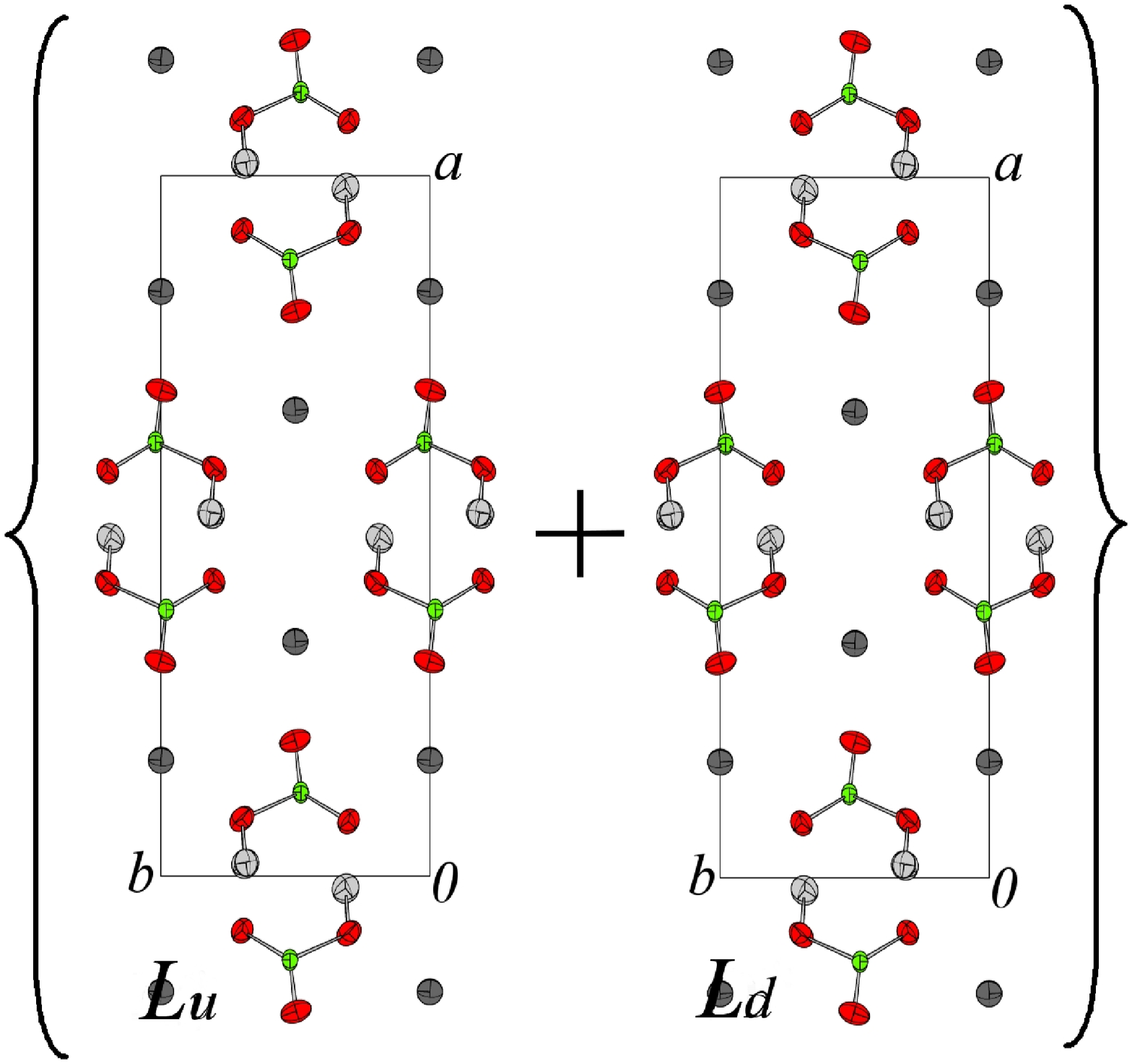}
\includegraphics[angle=0.,scale=0.17]{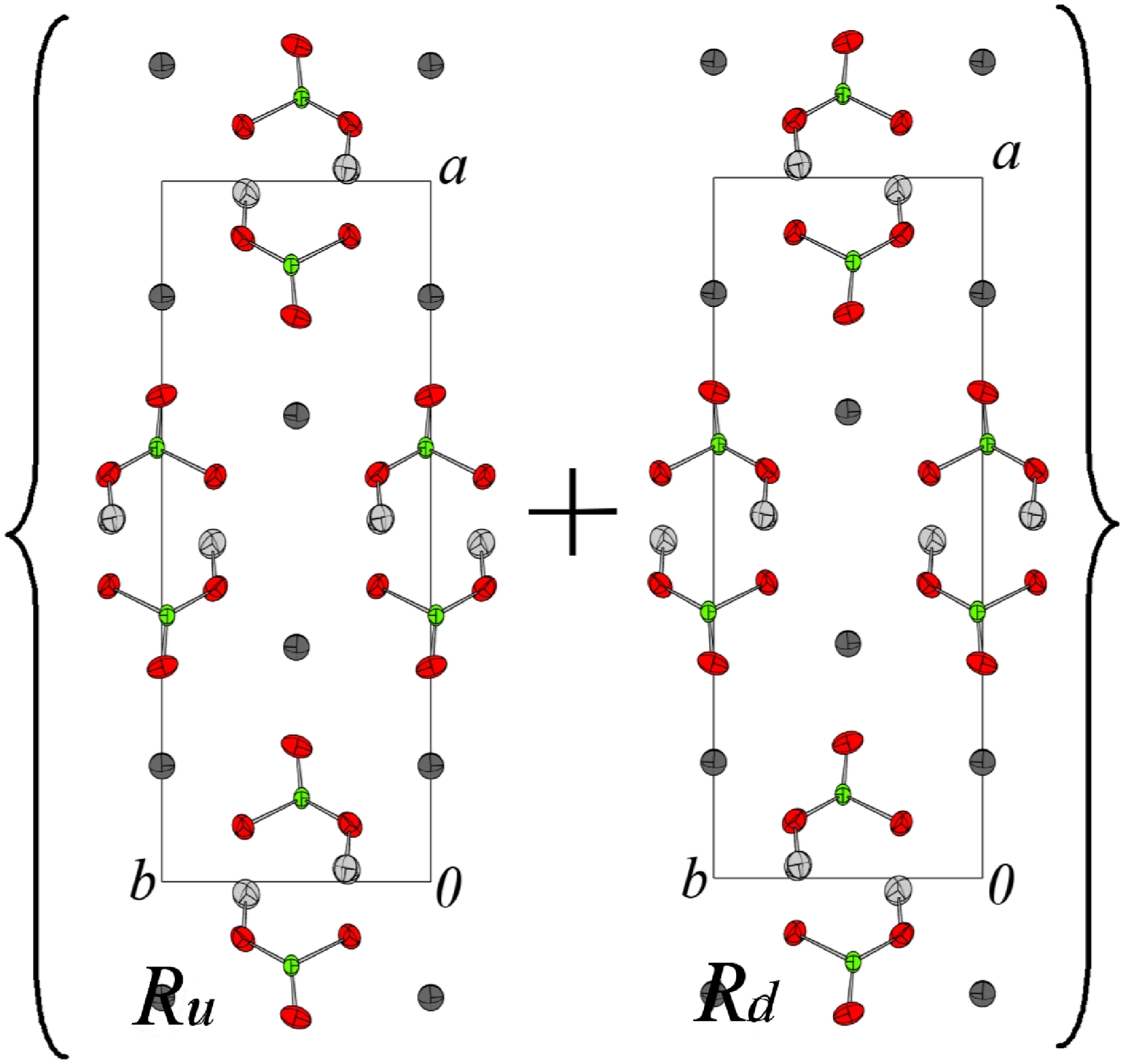}
\caption{\label{fig:4} Schematic representation of the $L_u$, $L_d$, and $R_u$, $R_d$, $P2_1/a$-like configurations corresponding to state vectors superposed in the $|C2/m\rangle$ crystal state above $T_c$ (see text). $L$ and $R$ correspond to different configurations for protons in dimers. Configurations $u$ and $d$ are symmetric with respect to ($a,c$). Each site is half-occupied.}
\end{center}
\end{figure}

For each substructure, there are two configurations, $L$ or $R$, for the proton sublattice. Within the framework of the adiabatic separation, the state vectors of the four configurations $L_u$, $R_u$, $L_d$, $R_d$, sketched in Fig. \ref{fig:4}, are $|p_u\rangle |H_{Lu\tau}\rangle |\tau\rangle$, $|p_u\rangle |H_{Ru\tau}\rangle |\tau\rangle$, $|p_d\rangle |H_{Ld\tau}\rangle |\tau\rangle$, $|p_d\rangle |H_{Rd\tau}\rangle |\tau\rangle$, where $|p_u\rangle$ and $|p_d\rangle$ represent heavy atoms. The ground state is a superposition of the degenerate states $|p_u\rangle |H_{Lu\tau}\rangle |\tau\rangle$ and $|p_d\rangle |H_{Ld\tau}\rangle |\tau\rangle$, while $|p_u\rangle |H_{Ru\tau}\rangle |\tau\rangle$ and $|p_d\rangle |H_{Rd\tau}\rangle |\tau\rangle$, also degenerate, are at higher energy ($\Delta E_{LR}$). The state vector of the crystal can be then written as 
\begin{equation}\label{eq:11}
\begin{array}{lcl}
|C2/m\rangle & = & 2^{-1/2} \{ \alpha(T) [|p_u\rangle |H_{Lu\tau}\rangle + |p_d\rangle |H_{Ld\tau}\rangle ] \\
& + & \beta(T) [|p_u\rangle |H_{Ru\tau}\rangle + |p_d\rangle |H_{Rd\tau}\rangle \} |\tau\rangle, 
\end{array}
\end{equation}
with $\alpha(T)^2 +\beta(T)^2 = 1$. The superposition of proton states is consistent with enhanced diffraction because the corresponding sublattices are identical in reciprocal space. 

QENS measurements of KHCO$_3$ \cite{EGS} are also consistent with (\ref{eq:11}). The inverse correlation time for proton transfer ($\tau_0^{-1} \approx 2\times 10^{12} s^{-1}$) is temperature independent, without any discontinuity at $T_c$, because QENS, like Raman spectroscopy, probes proton dynamics of the $P2_1/a$ structure below $T_c$ and the same dynamics for either $|P_u\rangle$ or $|P_d\rangle$ above $T_c$. 

Furthermore, $\tau_0^{-1}$ below $T_c$ is consistent with a superposition of proton states giving rise to quantum interferences whose beating frequency, $\nu_b \approx 2.5 \times 10^{12} s^{-1}$, is in reasonable agreement with QENS measurements \cite{Fil7}. Above $T_c$, quantum interferences for (\ref{eq:11}) are cancelled by symmetry. This is at odds with the expectation that fluctuations should increase at elevated temperatures. However, this fluctuation-free state cannot be ``observed'' because QENS probes either $|P_u\rangle$ or $|P_d\rangle$ states. 

Above $T_c$, diffuse scattering along $(a^*)$, very slow dynamics, and enhanced diffraction reported in previous works suggest dynamical interconversion between superposed $|P_u\rangle$ and $|P_d\rangle$ states, on the one hand, and their analogues $|P2_1/a\rangle_u$ and $|P2_1/a\rangle_d$ states, on the other. This may occur through very small displacements of heavy atoms along $(b)$ (Fig. \ref{fig:1}). Upon cooling, symmetry breaking at $T_c$ should lead to either $|P2_1/a\rangle_u$ or $|P2_1/a\rangle_d$ states, with equal probabilities. However, these structures are related through a translation vector $(a/2,b/2,0)$. They are indistinguishable and the symmetry breaking leads to a unique $P2_1/a$ structure. 

It is often argued that a complex open system undergoing continuous interactions with the environment should be in a significantly mixed state that cannot be represented by a state vector \cite{Vedral,JR1,JR2}. The present work leads to conclusions at variance with this view: the crystal of KHCO$_3$ at thermal equilibrium with the surrounding atmosphere can be represented by a state vector. The gaseous environment can be treated as a distribution of plane waves that cannot destroy the entanglement intrinsic to the lattice periodicity. Only the spin-symmetry can be transitory destroyed. However, this reversible disentanglement does not increase with time and it is insignificant because the density-of-states of the surroundings is smaller than that of the crystal by many orders of magnitude. 

\section{\label{sec:7}Conclusion}

We confirm with single-crystal neutron diffraction the $C2/m$ structure composed of split $C_i$ dimers proposed by Kashida and Yamamoto \cite{KY}, but proton positions are at variance with those based on X-ray diffraction. The structure of the sublattice of protons is practically temperature independent, apart from site occupancies. The enhanced diffraction pattern measured at 340 K is basically identical to that arising from macroscopic quantum correlations below $T_c$ \cite{FCKeen,FCG2}. 

The theoretical framework for quantum correlations is based on fundamental principles of quantum mechanics. The fermionic nature of protons, unveiled by the adiabatic separation, leads to spin-symmetry and super-rigidity. The spin-symmetry is probed by neutrons diffracted at $\mathbf{Q}$-values corresponding to nodes of the reciprocal lattice of protons. The coherent scattering cross-section is increased by a factor of $\approx 45$, compared to that for uncorrelated protons, and super-rigidity cancels the shortfall of intensity by the Debye-Waller factor. 

Above $T_c$, statistical disorder, or stochastic proton jumps, or clusters, or domains, or uncorrelated dynamics of individual dimers, or whatever phenomena which could destroy the translational invariance of the crystal, are excluded. The phase transition occurs from one ordered $P2_1/a$ structure to another ordered $C2/m$ structure. Both phases can be treated as pure states represented by state vectors. Raman spectroscopy and QENS show that eigen states for protons are identical in the two structural phases. This suggests that the state vector $|C2/m\rangle$ is a superpositions of state vectors $|P_u\rangle$ and $|P_d\rangle$ for $P2_1/a$-like structures. This leads to the rather paradoxical conclusion that quantum interferences arising from superposition of proton states are cancelled by symmetry, as long as these states are not ``observed''. 

This work emphasizes that atoms in the crystal-field of KHCO$_3$ are not individual particles possessing properties on their own right. They merge into macroscopic states and exhibit all features of quantum mechanics: nonlocality, entanglement, spin-symmetry, superposition and interferences. There is every reason to suppose that similar quantum effects should occur in many hydrogen bonded crystals undergoing structural phase transitions. 

\ack We should like to thank F. Romain from LADIR-CNRS for recording the Raman spectra. 

\section*{References}

\bibliographystyle{unsrt}

\begin{table}[!hbtp]
\caption{\label{tab:1} Neutron single crystal diffraction data and structure refinement for potassium hydrogen carbonate at various temperatures. $\lambda$ = 0.8305 \AA, space group $P 2_1/a$ at 300 K (after Ref. \cite{FCG2}) and $C 2/m$ at 340 K (this work). The dimer symmetry is either $C_i$ or $C_{2h}$ in the two different structure models under consideration. The variance for the last
digit is given in parentheses.}
\begin{indented}
\item[]\begin{tabular}{llll}
\br
 Crystal data  & 300 K & 340 K & 340 K\\
&  & $C_i$ & $C_{2h}$ \\
\mr
$a$(\AA) & 15.18(1) & 15.173(9) & 15.173(9) \\
$b$(\AA) & 5.620(4) & 5.628(4) & 5.628(4) \\
$c$(\AA) & 3.710(4) & 3.711(4) & 3.711(4) \\
$\beta$  & 104.67(5)$^\circ$ & 104.63(5)$^\circ$ & 104.63(5)$^\circ$ \\
Volume   & 306.2 & 306.6 & 306.6 \\
Reflections measured & 1731 & 1789 & 1789 \\
Independent reflections & 1475 & 931 & 931 \\
Reflections used      & 1068 & 752 & 755 \\
$\sigma$(I) limit & 3.00 & 3.00 & 3.00\\
Refinement on F & \\
R-factor & 0.035 & 0.028 & 0.038 \\
Weighted R-factor & 0.034 & 0.019 & 0.037 \\
Number of parameters & 56 & 59 & 34 \\
Goodness of fit & 1.025 & 1.116 & 1.107 \\
Extinction & 3260(100) & 15.4(4) & 14.9(7)\\
\br
\end{tabular}
\end{indented}
\end{table}

\begin{table}[!hbtp]
\caption{\label{tab:2} Atomic positions, isotropic temperature factors and site occupancies for KHCO$_3$ at 340 K. The variance for the last digit is given in parentheses. }
\begin{indented}
\item[]\begin{tabular}{llllll}
\br
 Atom & $x/a$ & $y/b$ & $z/c$ & U(iso)(\AA$^2$) & Occupation\\
\mr
K(1) & 0.16526(7) & 0.00       & 0.2957(3)  & 0.0290  & 1.00 \\
C(1) & 0.11951(3)  & 0.5192(1)  & $-0.1437(1)$ & 0.0186 & 0.50  \\
O(1) & 0.19339(4)  & 0.5019(1)   & 0.0950(2)   & 0.0354 & 0.50  \\
O(2) & 0.0781(2)   & 0.3021(1)  & $-0.264(1)$ & 0.0337 & 0.50  \\
O(3) & 0.0815(2)   & 0.6989(2)  & $-0.280(1)$ & 0.0301 & 0.50 \\
H(10) & 0.0176(2)  & 0.687(2)    & $-0.439(1)$    & 0.0462 & 0.25 \\
H(11) & 0.0189(2)  & 0.684(1)    & $-0.450(1)$    & 0.0312 & 0.25  \\
\br
\end{tabular}
\end{indented}
\end{table}

\begin{table}[!hbtp]
\caption{\label{tab:3} Thermal parameters in \AA$^2$ units for KHCO$_3$ at
340 K. The variance for the last digit is given in parentheses. The thermal parameters account for the variation of the contribution of each atom to Bragg's peak intensities through the thermal factor $T^{at}$ depending on the reciprocal lattice parameters $a^*$, $b^*$, $c^*$, and unit cell indexes in reciprocal space $h, k, l$, as $T^{at} = \exp [ -2\pi^2 (U_{11}^{at}h^2a^{*2} + U_{22}^{at}k^2b^{*2} + U_{33}^{at}l^2c^{*2}  + 2U_{12}^{at}ha^*kb^* + 2U_{23}^{at}kb^*lc^* + U_{31}^{at}lc^*ha^* ) ].$}
\begin{indented}
\item[]\begin{tabular}{lllllll}
\br
 Atom & U$_{11}$ & U$_{22}$ & U$_{33}$ & U$_{23}$ & U$_{13}$ & U$_{12}$ \\
\mr
K(1)  & 0.0316(4) & 0.0303(4) & 0.0227(3) & 0.000      & 0.0025(3) & -0.0003(1) \\
C(1) & 0.0209(2) & 0.0117(3)& 0.0216(2) & 0.0003(2) & 0.0023(1)& -0.0001(2) \\ 
O(1)  & 0.0247(2) & 0.0441(3)& 0.0315(3) & -0.006(3) & -0.0036(2)& 0.0083(1) \\
O(2)  & 0.029(1)& 0.023(1)& 0.041(2)& -0.0001(8) & -0.0067(9) & -0.0043(7) \\
O(3)  & 0.030(1) & 0.028(1) & 0.0295(9) & 0.0018(7) & 0.0009(7) & -0.0072(8) \\
H(10)  & 0.040(6) & 0.034(5) & 0.055(6)  & -0.003(5)  & -0.007(4)   & 0.003(4) \\
H(11)  & 0.034(5) & 0.023(4) & 0.036(5)  & 0.004(3)  & -0.008(5)   & -0.001(3)   \\
\mr
\end{tabular}
\end{indented}
\end{table}

\begin{table}[!hbtp]
\caption{\label{tab:4} Interatomic distances in \AA\ units in KHCO$_3$ at 340 K. The variance for the last digit is given in parentheses. }
\begin{indented}
\item[]\begin{tabular}{llll}
\br
C(1)---O(1) & $\left\{\begin{tabular}{l} 1.245(1)\\1.247(1)\\ \end{tabular} \right.$ & O(3)---H(10) & \ \ \ 1.002(1)\\
C(1)---O(2) & $\left\{\begin{tabular}{l} 1.395(1)\\1.210(1)\\ \end{tabular} \right.$ & O(3)---H(11) & \ \ \ 1.003(1) \\
C(1)---O(3) & $\left\{\begin{tabular}{l} 1.210(1)\\1.395(1)\\ \end{tabular} \right.$ & H(10)---H(11) & $\left\{\begin{tabular}{l} 0.053(3) \\0.601(1) \\ \end{tabular} \right.$ \\
C(1)---C(1) & \ \ \ 0.216(1) & H(10)---H(10) & \ \ \ 0.607(1) \\
O(1)---O(1) & \ \ \ 0.02(1) & H(11)---H(11) & \ \ \ 0.599(1) \\
O(2)---O(3) & \ \ \ 0.09(1) &  & \\
O(2)---H(10) & \ \ \ 0.983(1)   & O(2)---H(11) & \ \ \ 0.988(1) \\
\br
\end{tabular}
\end{indented}
\end{table}

\begin{table}[!hbtp]
\caption{\label{tab:5} Orders $n_y$, $n_x$ and positions $Q_y$, $Q_x$ of rods of intensity arising from the entangled array of orthogonal doubles lines of protons in two dimensions. Obs. $Q_x$: positions in \A\ of the observed rods of intensity along $Q_z$ in Figs \ref{fig:2} and \ref{fig:3}. $Q_yD_y/\pi$ is rounded to integers.}
\begin{indented}
\item[]\begin{tabular}{lllllll}
\br
 $n_y$ & $\tau_i\tau_f$ & $Q_y$ (\A)& $Q_yD_y/\pi$ & k = $Q_y/b^*$ & $Q_x$ & Obs. $Q_x$ (\A) \\
\mr
0 & $+1$ & 0    & 0    & 0 & $n_x\pi/x_0$ & $0,\pm 10$\\
1 & $-1$ & 2.86 & 5  & 2.57 & $(n_x+1/2)\pi/x_0$ & $\pm 5, \pm 15$\\
2 & $+1$ & 5.71 & 10 & 5.14 & $n_x\pi/x_0$ & $0,\pm 10$\\
3 & $-1$ & 8.57 & 15 & 7.71 & $(n_x+1/2)\pi/x_0$ & $\pm 5, \pm 15$\\
\br
\end{tabular}
\end{indented}
\end{table}

\end{document}